# Materials Research Express

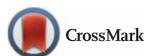

PAPER

OPEN ACCESS

# Industrial manufacturing and characterization of multiscale CFRP laminates made from prepregs containing graphene-related materials

Verónica Rodríguez-García[1,2], Julio Gómez[3], Francesco Cristiano[4] and María R Gude[1]

[1] FIDAMC, Foundation for the Research, Development and Application of Composite Materials, Avda. Rita Levi Montalcini 29, 28906 Getafe, Madrid, Spain
[2] Departamento de Ciencia de los Materiales, ETSI Caminos, Canales y Puertos, Universidad Politécnica de Madrid C/Profesor Aranguren s/n, 28040 Madrid, España
[3] Avanzare Innovacion Tecnologica S.L. Avda. Lentiscares 4-6. 26370 Navarrete, Spain
[4] Nanesa S.r.l Via del Gavardello, 59C, Arezzo, Tosc. 52100, Italy

E-mail: maria.r.rodriguez@fidamc.es

## Abstract

The introduction of graphene-related materials (GRMs) in carbon fibre-reinforced polymers (CFRP) has been proved to enhance their mechanical and electrical properties. However, methodologies to produce the 3-phase materials (multiscale composites) at an industrial scale and in an efficient manner are still lacking. In this paper, multiscale CFRP composites containing different GRMs have been manufactured following standard procedures currently used in the aerospace industry with the aim to evaluate its potential application. Graphite nanoplateletetelets (GNPs), *in situ* exfoliated graphene oxide (GO) and reduced graphene oxide (rGO) have been dispersed into an epoxy resin to subsequently impregnate aeronautical grade carbon fibre tape. The resulting prepregs have been used for manufacturing laminates by hand lay-up and autoclave curing at 180 °C. A broad characterization campaign has been carried out to understand the behaviour of the different multiscale laminates manufactured. The degree of cure, glass transition temperature and degradation temperature have been evaluated by thermal evolution techniques. Similarly, their mechanical properties (tensile, flexural, in-plane shear, interlaminar shear and mode I interlaminar fracture toughness) have been analysed together with their electrical conductivity. The manufacturing process resulted appropriated for producing three-phase laminates and their quality was as good as in conventional CFRPs. The addition of GO and rGO resulted in an enhancement of the in-plane shear properties and delamination resistance while the addition of GNP improved the electrical conductivity.

## 1. Introduction

Carbon fibre-reinforced polymers (CFRPs) are widely used in the aerospace industry due to their outstanding strength-to-weight ratios [1, 2]. Nevertheless, the fact that CFRPs are made by staking fibre plies and held by a resin matrix makes these materials highly anisotropic, showing poor conductivity and poor mechanical properties in the direction perpendicular to the fibres, leading to delamination, brittle failure and poor damage resistance [3, 4]. The scientific community has proposed several techniques to enhance this behaviour such as z-pinning/stitching, fibre sizing or matrix modifications [4–6] and interleaving [7]. The use of thermoplastic non-woven veils doped or coated with conductive nanoparticles can increase their fracture toughness [7–10] while providing multifunctionality [7, 8, 11], but they have to be positioned by hand, not being compatible with the automated manufacturing processes. Up to the date, there is a lack of strategies that are adequate for producing enhanced CFRPs at an industrial scale without increasing largely the costs or production times, thus making them inefficient and compromising its potential application [6].

Among the different strategies, the addition of micro- and nanoreinforcements to the resin matrix has also demonstrated to enhance the polymer and interplay characteristics [4, 12–14]. Particularly, graphene-related materials (GRMs) are promising candidates to act as nanoreinforcements due to their outstanding properties inherited from graphene [15–17]. The strength, stiffness and toughness of GRMs, as well as their thermal and electrical conductivities, together with the large specific surface area and aspect ratio, lead to an enhancement of the mechanical properties of polymers while providing multifunctionality [15–19]. Furthermore, the functionalization and surface modification of graphene have been recognized as an effective strategy to improve the interface properties between the GRMs and epoxy matrices [19–23]. GRMs can be prepared at low temperature, without metal catalysis, by exfoliation of bulk graphite with ultrasonic methods, high speed mixing or water-jet milling [15], unlike other nanocarbon reinforcements as fullerenes and carbon nanotubes (CNTs). Actually, the exfoliation mechanism allows producing large quantities and it is a scalable method suitable at industrial scale [24, 25]. In addition, while one of the main problems of adding nanocarbon reinforcements to the resin is the subsequent increase in viscosity, it has been demonstrated that this increment is lower using GRMs than CNTs, what is a key aspect to manufacture multiscale composites (3-phase composites) in a large scale [17, 26–28].

Several routes are available for the incorporation of the GRMs into the matrix together with the carbon fibres (CFs) [6, 15], but these routes are limited when looking for methods applicable in the aerospace industry. Resin transfer moulding (RTM) or vacuum assisted resin infusion molding (VARIM) have been investigated to produce multiscale laminates [29–32], but these methods have shown limitations related to GRM agglomeration, filtration effects and high viscosity of the nanoreinforced resins [6, 29–31]. An alternative methodology prepreg lay-up and autoclave, which is curing is a well-established method for producing high performance composites with well consolidated manufacturing routes [33]. There are different alternatives for obtain three-phase composites using prepregs: (1) pre-impregnate fabrics or fibre tapes with a previously nanoreinforced resin to produce 'doped' prepregs, (2) insertion of nanofiller layers between conventional prepregs and (3) modification of conventional prepregs by nanofiller spraying [6, 33]. Among them, the first method is the most compatible with the current automated lay-up technologies. However, the number of studies of multiscale CFRPs containing carbon nanofillers manufactured by prepreg lay-up and autoclave curing scarce in the literature and no study following industrial procedures has been reported to the knowledge of these authors.

Siddiqui *et al* produced prepregs with carbon nanotubes (CNTs) with a laboratory-scale prepregger [34] and observed an increase in the interlaminar shear strength by 12% and in the torsional shear modulus and strength (17% and 19.5% respectively) when adding 0.5% of CNTs and curing with a vacuum hot press at 120 °C [35]. Joshi *et al* sprayed multiwalled CNTs on woven CFRP prepreg and cured the laminates in a vacuum oven at 120 °C for the enhancement of the interlaminar properties [36]. Yokozeki *et al* used cup-stacked CNTs to charge an epoxy resin and develop prepregs that were cured in an autoclave at 130 °C, obtaining improvements in interlaminar fracture toughness [37]. Wang *et al* produced prepregs by hot press moulding with CF fabric and manufactured composites by hot press moulding (150 °C) and post-curing in an oven (140 °C), reporting improvements in flexural, tensile and fracture toughness properties of the CFRP by incorporating a combination of GNPs and CNTs at 1 wt.% [38]. Finally, Mannov *et al* manufactured prepregs containing thermally reduced graphene oxide with a filament winding machine and cured them in an autoclave at 80 °C (with a post-curing at 140 °C), obtaining laminates with an enhancement of the impact damage and higher residual compressive properties [39]. Those studies probe that is possible to improve the mechanical properties by the addition of GRMs when using prepreg technology, however they have been carried out at a laboratory scale.

Great efforts from the European Community are being invested in developing scalable procedures in science to transfer them to the industry [6, 15, 40, 41]. Within this framework, and considering the potential industrial benefits of the introduction of GRM in CFRPs for the aerospace industry, this study aims to produce multiscale CFRPs laminates with commercial materials, involving industrial partners and using industrial procedures in every step of the process: from the production of GRMs to the final manufacturing of multiscale laminates. To that end, three kinds of commercial GRMs have been selected as nanoreinforcements with the aim of evaluating its potential application: graphite nanoplatelets (GNPs), graphene oxide (GO) obtained from *in situ* exfoliation of graphite oxide and reduced graphene oxide (rGO). A 2 wt.% of the GRMs have been mechanically dispersed in a solventless process in an epoxy resin and subsequently carbon fibre-reinforced epoxy prepregs (aerospace grade) have been produced by a hot melt process which, in turn, has been used to fabricate laminates cured in an autoclave, obtaining three different kinds of multiscale laminates. A complete characterization of the multiscale laminates is presented to evaluate the potential application and enhancements of the laminates produced. To the best of our knowledge, this is the first scientific paper exploring the use of GRM-containing prepreg to produce multiscale laminates using procedures already implemented in the industry.

## 2. Materials and methods

### 2.1. Materials

The matrix used for manufacturing the prepregs consists of a multifunctional basis Bisphenol A diglycidyl ether (DGEBA) epoxy system (1.22 g cm$^{-3}$) supplied by Delta-tech (Rifoglieto 60/a - int.1 55011 Altopascio, Italy) with a curing temperature of 180 °C. This resin, named EM180, was used in Elmarakbi *et al* work [42] (named as 'resin B'). Unidirectional carbon fibre tape of aerospace grade T700G-12K (1.80 g cm$^{-3}$ and FAW 200 gsm) provided by Toray Industries was chosen as the reinforcement.

Three different kinds of GRMs were selected as nanoreinforcements:

1. GNPs, commercial graphite nanoplatelets produced by Nanesa (G2Nan), obtained by exfoliation of graphite. This material was used as nanoreinforcement as well by Elmarakbi *et al* [42].

2. GO (graphene oxide) was prepared by modified Hummers' method from graphite flakes [43] by Avanzare for *in situ* exfoliation in the resin.

3. rGO (reduced graphene oxide) was obtained from the thermochemical reduction of the previous GO by Avanzare [43].

Details of the different GRMs provided by the suppliers are given in table 1. SEM and TEM images of the three GRMs are given in the Supplementary Information (SI) (figure S1) which is available online at stacks.iop.org/MRX/7/075601/mmedia. A description of the techniques used for the characterization presented in table 1 can be found as well in the SI.

The number of layers of the graphene materials ($N_G$) was calculated by dividing the maximum theoretical SSA of graphene by the experimentally determined BET SSA ($N_G = 2630/BET$) [42, 44]. The BET SSA of the materials (table 1) is far below the theoretical value of fully exfoliated pristine graphene. Considering this factor, the $N_G$ for GNP is higher than the value observed for GO and rGO. In addition, XPS analysis shows 0.9% of oxygen for GNP, 30.6% for GO and 7.9% for rGO. This indicates that the graphite nanoplatelets are pristine and could allow obtaining good electrical properties. GO and rGO, in turn, could present stronger interfacial interactions with the resin due to their high oxygen content.

### 2.2. Addition of GRMs to the resin

The addition of the GRMs to the resin was performed by the GRMs suppliers who optimized the processes to obtain a good dispersion of them. Therefore, different processes were followed for the addition of GNPs (by Nanesa) and for the GO and rGO (by Avanzare) to the resin.

For the mixing of the GNPs, the neat resin was heated at 60 °C for 2 h to reduce viscosity up to 49 Pa·s [42]. Then, the GNPs were added slowly and gradually while being dispersed with Disperlux cowless system (shaft mixing) at 2000 rpm for 60 min. Then, there was a second stage of homogenization and distribution of the nanoparticles with a Silverson high shear batch mixer at 3000 rpm for 30 min. This system allows homogenizing the GNPs by breaking the agglomerates formed during the initial dry mixing.

The GO was dispersed in the resin using a DISPERMAT at 13.000 rpm with a cowless helix for 20 min (10 min + 10 min to avoid the overheating of the motor due to the high viscosity of the resin). Then, a dip ultrasonication was applied for 1 h by a H40 sonotrode (400 W) for *in situ* exfoliation [42]. The rGO, previously exfoliated, was dispersed into the resin using the same method as for the GO.

In this study the content of each GRM added to the resin is 2% by weight. In all the cases, viscosity was evaluated to ensure that it was appropriate for the production of prepregs according to Delta-tech requirements. The resin viscosity measured at 60° degrees (pre-impregnation process temperature) showed viscosity increments of 138, 26 and 18% when adding GNP, GO and rGO, respectively.

### 2.3. Epoxy-GRM composites production and characterization

Once the GRMs were added to the base of the epoxy resin, a small portion of the resin was used for the evaluation of the GRMs dispersion. The formulation of these resins was finished by adding the catalyst and the hardener and mixing. After completion of the formulation, the samples were cast and cured 2 h at 180 °C, with heating rates of 1 °C min$^{-1}$. For the SEM characterization, epoxy-GRM composites were cryofractured and were studied by SEM using a Hitachi S-2400 microscope.

**Table 1.** Main characteristics of the GRMs present in this study. Details about the characterization techniques can be found in the Supplementary Information (SI).

|  | GNPs | GO | rGO |
|---|---|---|---|
| GRM type | Nanoplatelets | Graphene oxide | Reduced graphene oxide |
| Average particle lateral size ($\mu$m)[a] | 25 | 43 | 39 |
| Average particle lateral size ($\mu$m)[b] | 15–30 | 20–25 | 20–25 |
| BET (m$^2$ g$^{-1}$) | 30 | 562 | 780 |
| Average N° of layers ($N_G$)[c] | 87.7 | 4.7 | 3.4 |
| Average Flake thickness (nm)[d] | 14 | 3 | 1 |
| % Ox | 0.9 | 30.6[e] | 12[f] |

[a] Measured using laser diffraction in solid D50.
[b] Measured by SEM.
[c] $N_G$ = 2630/BET.
[d] Measured by TEM.
[e] Measured by XPS after dry: (C=O: 23%, C–O: 77%).
[f] Measured by XPS after dry (C=O: 35%, C–O: 43%).

### 2.4. Prepreg and laminate production

Once the GRM was added to the base of the epoxy resin, the formulation of the resin was finished adding the catalyst and the hardener and mixing at 60 °C using a laboratory mixer under vacuum. This mixture was filmed on silicon paper at the same temperature. Then, the fibres were impregnated by a hot melt process at 65 °C [42], resulting in prepregs containing a 34% by weight of resin, which means a 0.68% by weight of the corresponding GRM in the prepreg and a ply thickness of 0.2 mm. An additional batch was made with neat resin, to be used as a control.

The laminate production was performed by FIDAMC. Several laminates with different stacking sequences were manufactured by hand lay-up of the prepregs with vacuum compaction (figure S2). All the laminates were cured in an autoclave with vacuum bag (figure S3) at 6–7 bars. The heating rate applied was 1 °C min$^{-1}$ until reaching 180 °C and maintained for 120 min. The cooling rate was 2 °C min$^{-1}$ (corresponding autoclave cycle record in figure S4). Details about the different laminates configurations produced for specimen extraction can be observed in the SI. The different laminates have been identified as CFRP for the reference, CFRP-GNP, CFRP-GO and CFRP-rGO for the laminates having graphite nanoplatelets, graphene oxide and reduced graphene oxide, respectively.

A total of 16 laminates were manufactured and inspected using ultrasound technique. All the laminates reached an energy loss lower than 6 dB, showing no void or delamination (figure S5). Details about this technique are given in the SI.

### 2.5. Characterization of the laminates

#### 2.5.1. Characterization of the GRMs dispersion in the laminates

As seen, several steps are performed for producing the prepregs and laminates containing GRMs. It must be considered that the dispersion of the GRMs is presumably modified during the last step of the prepregs production, the pre-impregnation process by hot melt process. Because of this, a morphological evaluation of each laminate manufactured was performed. Three samples (20 × 20 mm$^2$) were obtained from all the laminates to evaluate the morphology and defects in polished cross sections with an optical microscope Nikon Eclipse LV150.

Similarly, to study the final dispersion of GRMs in the laminates, SEM images of the fracture surfaces of the coupons subjected to ±45° tensile test have been performed by using a Hitachi S-2400 microscope.

#### 2.5.2. Physicochemical characterization of the laminates

The fibre, resin and void contents ($M_{resin}$, $V_{fibre}$, $V_{void}$) were determined by extracting the resin by sulphuric acid digestion according to EN2564:2017 method A. To do that, three samples (10 × 20 mm$^2$) of each laminate with a configuration $[0]_{10}$ were tested. The density of each sample was determined by the immersion method with an analytical balance Mettler Toledo XP205 Delta Range (ISO1183-1:2012 Method A).

The degree of cure ($\alpha$) was characterized by differential scanning calorimetry (DSC) with a TA Instruments DSC Q2000. Two samples of each laminate manufactured and the corresponding uncured prepregs were tested under non-isothermal conditions with a heating rate of 10 °C min$^{-1}$ under a constant flow of nitrogen of 50 ml min$^{-1}$, following ISO 11357-5:2013 standard.

The glass transition temperature ($T_g$) was determined by dynamic mechanical analysis (DMA) with a TA Instruments DMA Q800 (5 °C min$^{-1}$ from RT to 270 °C, 1 Hz, 15 $\mu$m) using a single cantilever clamp according to EN6032:2015 Method A. Three samples of 35 × 10 mm$^2$ and lay-up [0]$_{10}$ of each material were tested.

For studying the degradation temperature of the laminates, three samples with lay-up [0]$_{10}$ of each material were tested by thermogravimetric analysis (TGA) with a TGA Q50 from TA Instruments. The samples were heated under air at a heating rate of 10 °C min$^{-1}$ from room temperature to 1000 °C (ISO 11358-1:2014).

*2.5.3. Mechanical characterization of the laminates*

Static tensile tests (ISO 527-1:2012) were carried out with a universal testing machine MTS landmark 370.10 System and a load cell of 100 kN at speed 2 mm min$^{-1}$. A total of five samples (250 × 25 mm$^2$) of each material were tested with configuration [0]$_{10}$, with 150 mm between tabs. A longitudinal extensometer was used for measuring strains, and fibreglass tabs were bonded to the samples.

Flexural tests in the 0° direction (EN 2562:1997) were performed with a universal testing machine AllroundLine Z01OTH from Zwick and a load cell of 10 kN to five samples (100 × 10 mm$^2$) with lay-up [0]$_{10}$ of each material. Test speed was 5 mm min$^{-1}$, and distance between supports was 80 mm. A pre-load of 5 N was applied and Zwick deflectometer was positioned to measure deflections.

Interlaminar shear properties were evaluated using the short-beam three point bend test (EN 2563:1997) were performed as well in the mentioned Zwick system and load cell of 10 kN at speed 1 mm min$^{-1}$ to six samples (20 × 10 mm$^2$) of each material with configuration [0]$_{10}$. The distance between supports was calculated as five times the thickness.

In-plane shear properties were studied by ±45° tensile tests (EN 6031:1995) were performed with the MTS 370 machine mentioned and a load cell of 100 kN to six samples (230 × 25 mm$^2$) with lay-up [+45/−45]$_{2s}$ and distance between tabs of 130 mm. Longitudinal and transversal extensometers were used for measuring strains. Furthermore, scanning electron microscopy by using a Hitachi S-2400 microscope was performed to the fracture surfaces of in-plane shear coupons.

Mode I interlaminar fracture toughness tests (prEN 6033:1995) were performed with the Zwick machine and a load cell of 10 kN. A total of six samples (250 × 25 mm$^2$) with configuration [0$_9$/0$_9$] of each material were tested with a side clamped beam fix (SCBF) tool. A release film was positioned during the fabrication of the laminates to act as artificial crack. The thickness edge was painted in white, then with the help of a microscope the end of the insert (at ~40 mm from the beginning of the sample), the end of the pre-crack (at ~50 mm from the beginning of the sample) as well as the minimum final distance that the crack should reach (at 110 mm from the beginning of the sample) were marked. The pre-crack was generated at a speed of 5 mm min$^{-1}$, while the crack was induced at a speed of 10 mm min$^{-1}$. Finally, measurements of the final crack lengths generated were taken with the help of the microscope.

*2.5.4. Electrical characterization of the laminates*

The effect of the GRMs in the electrical conductivity has been studied. Three samples (100 × 20 mm$^2$) with configuration [0]$_{10}$ of every material and 2 mm of thickness have been subjected to electrical resistance tests along the in-plane direction 'X/Y' and six samples (40 × 40 mm$^2$) in the through-thickness direction 'Z'. The corresponding opposite faces - edges for 'X/Y' direction and surfaces for 'Z' direction - of the specimens were coated with a ®Pelco Silver Paint from Ted Pella Inc. to ensure good contact between them and the probes of the multimeter. The test consists of injecting a direct electrical current into the specimen and measuring the voltage decrease between the surfaces of the specimens through a four-probe method with a Keithley 2410 ohmmeter from Keithley Instruments, Inc.

To ease the operational feasibility of the tests, specific jigs were used for supporting the specimens. The jigs present two brass side plates (electrodes) that ensure the electrical contact with the face of the specimens by adjusting them with a bolt at a constant torque.

For electrical measurements along the 'X/Y' direction, the Labtracer Sourcemeter 2.9 Integration Software of Keithley Instruments, Inc., has been used to apply a voltage (V) sweep of five points between 104–105 volts. On the other hand, for testing through the 'Z' direction, a sweep of intensity (I) of five points between 10 and 100 mA was applied.

## 3. Results and discussion

### 3.1. Epoxy-GRMs composites characterization

The GRMs agglomeration, dispersion and distribution in cryofracture samples of epoxy resin-GRMs composites have been observed by SEM (figures 1(a)–(c)). In all the cases, agglomerates from 10 to 20 $\mu$m in lateral size are observed. These agglomerates are composed of discrete particles of the GRMs stacked in the graphene plane. GNP agglomerates (figure 1(a)) shows a flat structure, however in the case of rGO a wavy structure is observed (figure 1(c)), attributed to the creation of defects during the oxidation and reduction process. In the case of GO composite, hollow spaces between different agglomerates and the resin are observed (figure 1(b)).

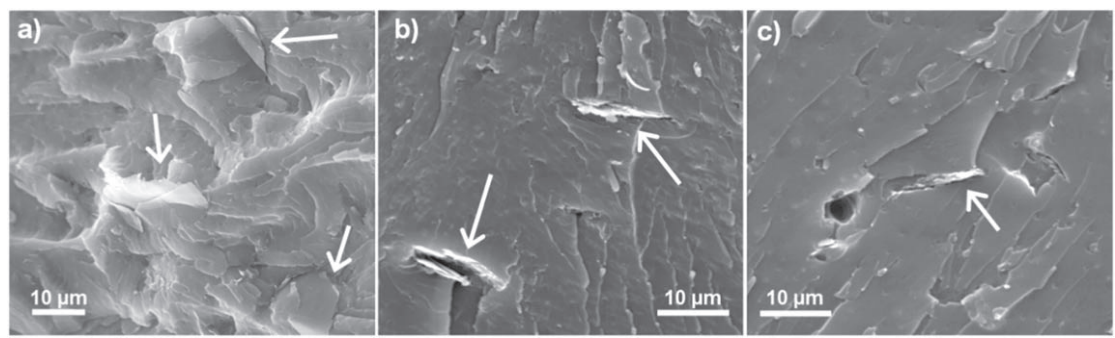

**Figure 1.** SEM micrographs of the epoxy-GRMs composites: (a) GNP at 2000×, (b) GO at 1500×, (c) rGO at 1500×. White arrows indicate the presence of the GRMs.

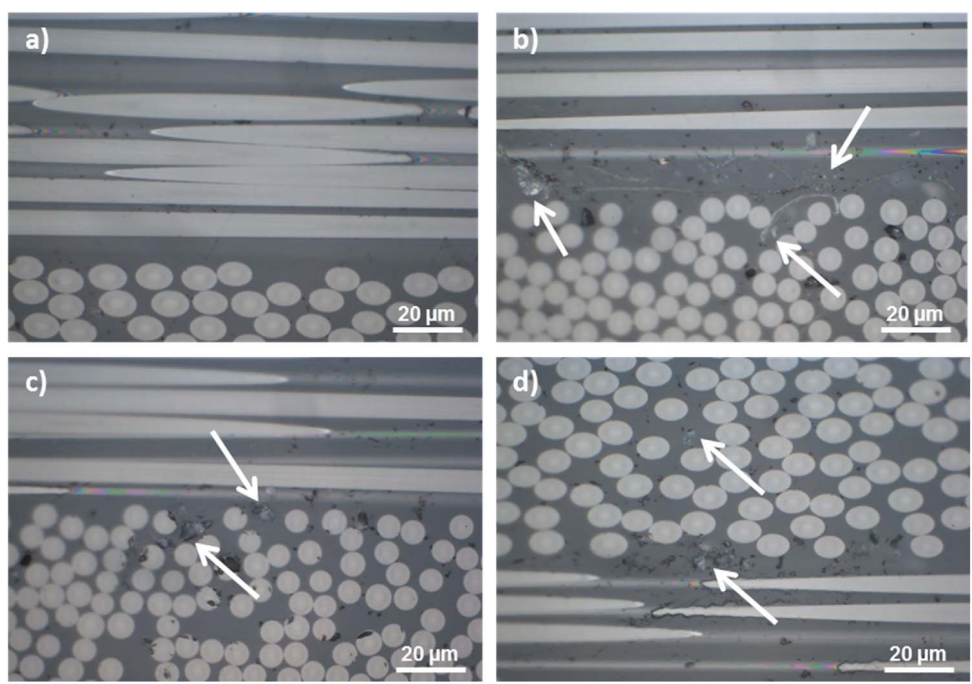

**Figure 2.** Optical micrographs at a magnification of 1000× of (a) CFRP laminate (b) CFRP-GNP laminate, (c) CFRP-GO laminate and (d) CFRP-rGO laminate. Arrows indicate GRM presence in the interlaminar region of the laminates.

### 3.2. Characterization of the laminates
*3.2.1. Characterization of the GRMs dispersion in the laminates*
The optical micrographs of the laminates show material integrity, parallel and well stacked layers in all the cases (figure S6). Details can be observed in figure 2, which shows optical micrographs at a high magnification (1000×). Some differences can be appreciated between the reference (figure 2(a)) and the multiscale laminates (figures 2(b)–(d)). In the multiscale materials, bright areas of 5 to 30 $\mu$m are observed in the interlaminar region (pointed with arrows) suggesting the presence of GRMs [30]. Although the average lateral size of the GO and rGO is larger than that of the GNPs (table 1), the particles observed in the micrographs are smaller in the CFRP-GO and CFRP-rGO than in CFRP-GNP laminates. This could be related to the presence of functional groups which interrupt the sp$^2$ carbon network leading to wrinkling effects when they are subjected to shear or in-plane compression, resulting in the deformation of the functionalized forms of GRMs [17].

In all the cases the observed particles are found in the interlaminar region and the first micrometres of the plies in the multiscale laminates, indicating that the size of the particles has hampered its penetration through the fibres. However, in figure 2(c) (CFRP-GO laminates), the particles are isolated and not big groups of particles have been found, while in figure 2(b) (CFRP-GNP laminates), a higher amount of particles concentrated in the interlaminar region can be observed, what may indicate agglomeration of GNPs. The particles in the CFRP-rGO laminate (figure 2(d)) seem to be in an intermediate state, seeing some small agglomerates. This difference in the dispersion of the nanoreinforcements among the samples could be due to the functional groups present in the GO and rGO. The hydroxyl groups present in GO and rGO (in less quantity as indicated in table 1) increase the electrostatic attractive interactions and hydrogen bonding with the epoxy, enhancing the dispersion and integration of the nanoparticles into the resin [17, 19, 45, 46]. On the other hand, the $\pi$–$\pi$ interactions and van der Waals forces in GNPs together with its higher size could prompt the formation of aggregates [28, 46], leading to GNP-rich and -poor regions [47].

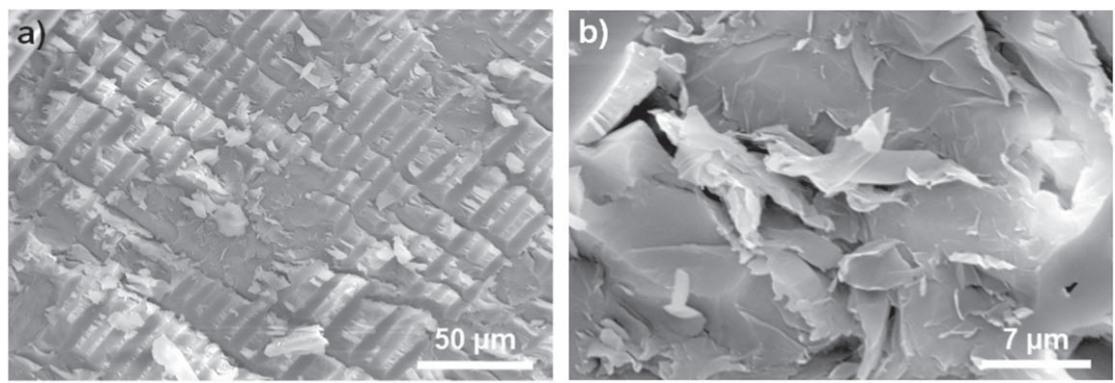

**Figure 3.** SEM images of the fracture surfaces of the coupons subjected to ±45° tensile test of CFRP-GNP laminate at (a) 600× and (b) 4000×.

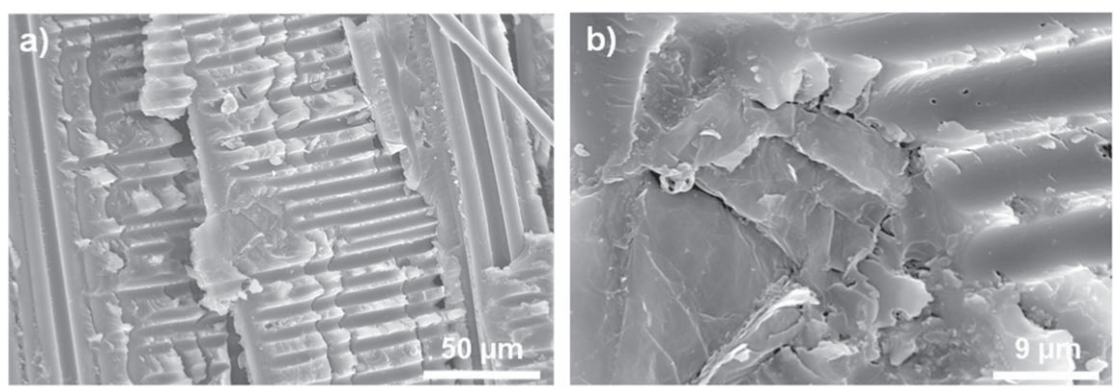

**Figure 4.** SEM images of the fracture surfaces of the coupons subjected to ±45° tensile test of CFRP-GO laminate at (a) 600× and (b) 3000×.

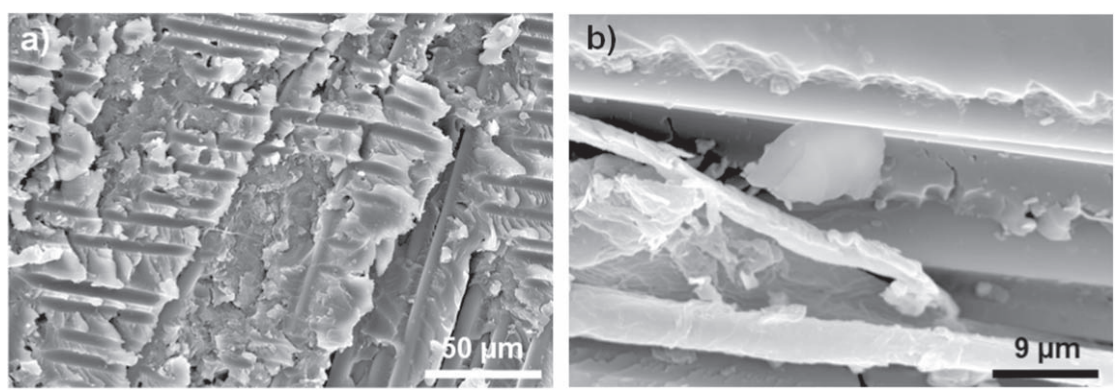

**Figure 5.** SEM images of the fracture surfaces of the coupons subjected to ±45° tensile test of CFRP-rGO laminate at (a) 600× and (b) 3000×.

**Table 2.** Results obtained in the physicochemical characterization of the multiscale and CFRP laminates: resin weight ($M_{resin}$), fibre volume ($V_{fibre}$) and void volume ($V_{void}$) contents, glass transition temperature ($T_{g\text{-}peak}$), decomposition temperature ($T_d$).

|  | $M_{resin}$ (%) | $V_{fibre}$ (%) | $V_{void}$ (%) | $T_{g\text{-}peak}$ (°C) | $T_d$ (°C) |
| --- | --- | --- | --- | --- | --- |
| CFRP | 33.7 ± 0.4 | 56.9 ± 0.4 | 0.41 ± 0.04 | 201.4 ± 0.9 | 353.8 ± 0.4 |
| CFRP-GNP | 34.6 ± 0.5 | 56.1 ± 0.6 | 0.14 ± 0.05 | 200.1 ± 0.3 | 353.0 ± 1.1 |
| CFRP-GO | 35.4 ± 2.0 | 55.2 ± 2.2 | 0.18 ± 0.02 | 200.9 ± 0.8 | 355.4 ± 1.4 |
| CFRP-rGO | 35.1 ± 1.1 | 55.5 ± 1.1 | 0.26 ± 0.12 | 201.1 ± 0.7 | 355.2 ± 1.3 |

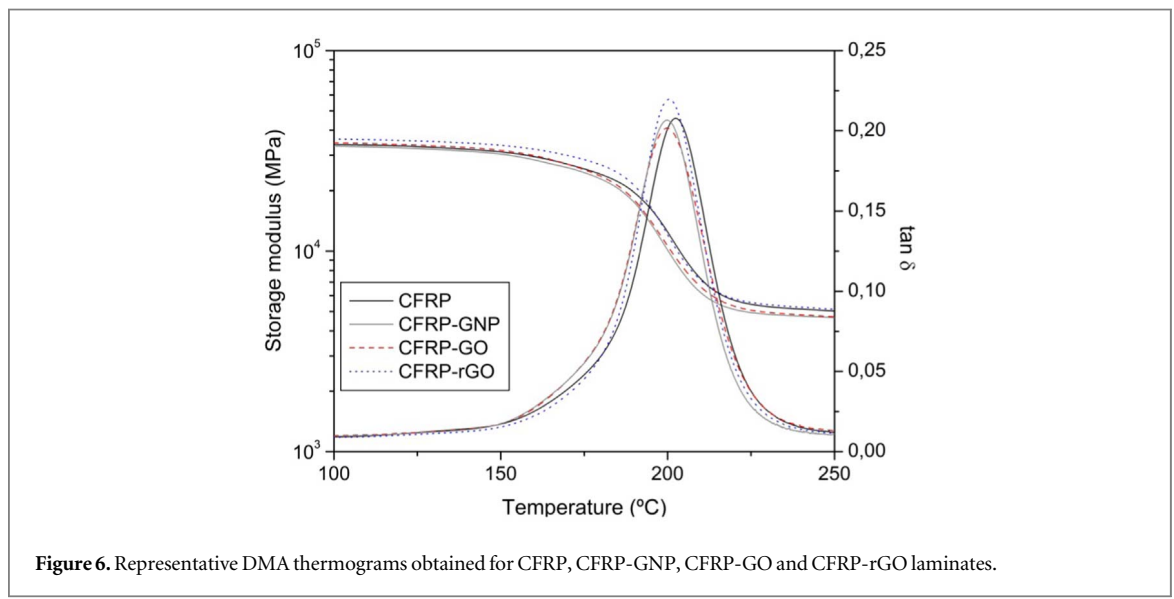

**Figure 6.** Representative DMA thermograms obtained for CFRP, CFRP-GNP, CFRP-GO and CFRP-rGO laminates.

In addition, fracture surfaces obtained from the broken in-plane shear coupons have been observed by SEM to study the GRMs dispersion in the multiscale laminates (figures 3–5 and S7(b)–(d)). The SEM images of CFRP-GNP broken coupon (figures 3 and S7(b)) show GNPs agglomerates and stacked discrete GNP particles, with sizes ranging from 10 to 30 $\mu$m between the plies. This would be in accordance with the previously observed micrographs of CFRP-GNP laminates (figure 2(b)). Regarding the CFRP-GO coupon, it should be noted that it was hard to find evidences of the nanoparticles and just one region with agglomerated GO particles was observed (figure 4). This failure region in the interplay shows similar stacking of GO particles (figure 4(b)) than the observed in the epoxy-GO composite (figure 1(b)). Finally, some rGO agglomerates could be observed in the CFRP-rGO broken coupon, however, they were found between fibres (figure 5(b)), pointing to higher penetration than in the case of the GNPs (figure S7(d), the arrows indicate graphene sheets). The rGO agglomeration observed in the interplay (figure 5(a)) could be similarly explained to the GNPs agglomeration, by the van de Waals forces and $\pi$–$\pi$ interactions between rGO nanoparticles, considering the restoration of the graphitic network of $sp^2$ bonds when reducing the GO [46].

*3.2.2. Physicochemical characterization of the laminates*
Details about the calculation method of resin weight ($M_{resin}$), fibre volume ($V_{fibre}$) and void volume ($V_{void}$) contents obtained from the acid digestion tests can be found in the SI. The results, shown in table 2, confirm that the fibre and resin contents are in accordance with the theoretical values (34% by weight of resin). Small variation in fibre content is observed from 56.9% in the case of unfilled resin to 55.2% in the GO filled one. The obtained void volume content is near 0%, in agreement with the non-destructive inspection and the microscopic analysis (figures 2 and S5 and S6) which showed no porosity (it should be noted that dark points found in the micrographs are caused by the polishing procedure but they do not indicate the presence of porosity in the laminates). These results indicate that it is possible to produce multiscale laminates with industrial quality through industrial approaches.

The laminates degree of cure was calculated using the average residual curing enthalpy and the total enthalpy of reaction from the corresponding uncured prepregs, as indicated in the standard. The results point to an effective cure cycle where all the laminates manufactured presented a degree of cure higher than 97%. Representative DSC thermograms of each material are presented in figure S8.

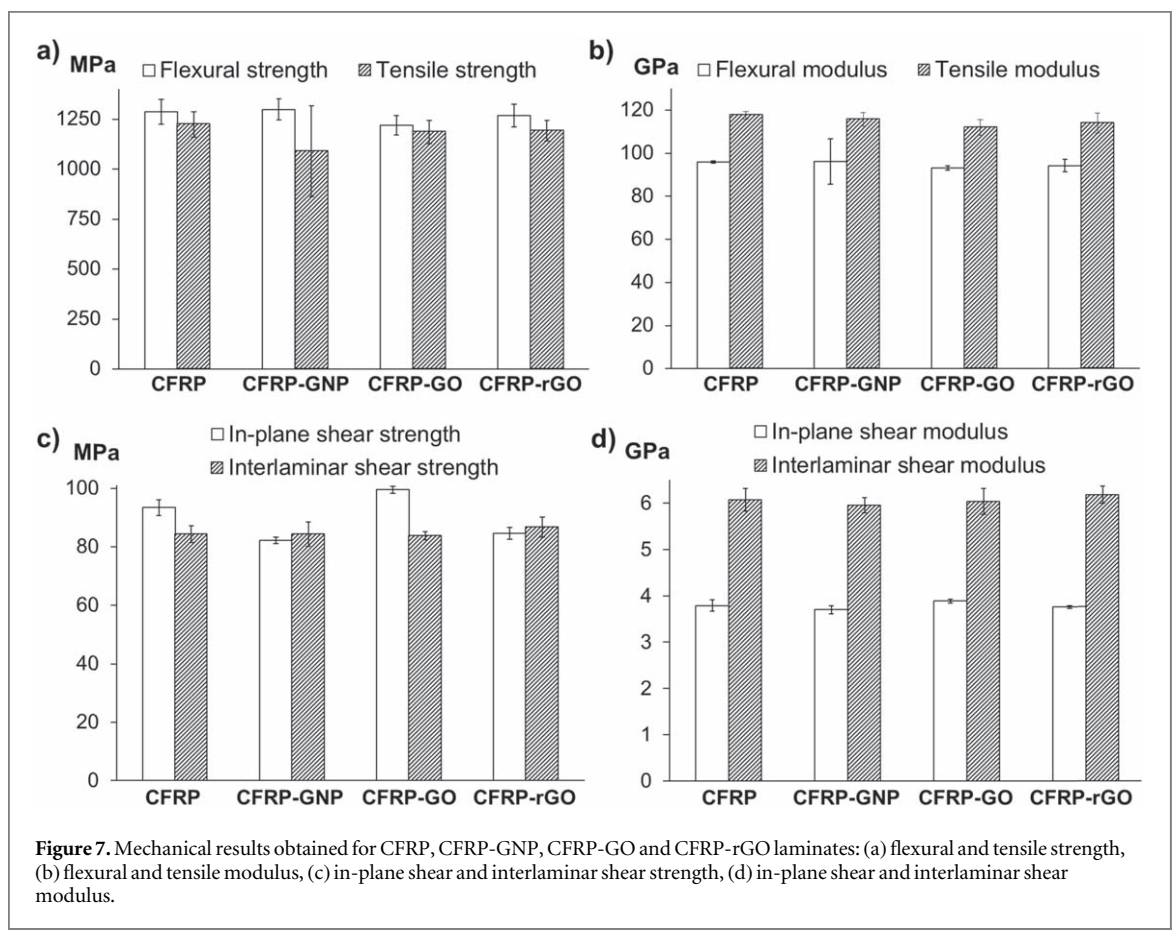

**Figure 7.** Mechanical results obtained for CFRP, CFRP-GNP, CFRP-GO and CFRP-rGO laminates: (a) flexural and tensile strength, (b) flexural and tensile modulus, (c) in-plane shear and interlaminar shear strength, (d) in-plane shear and interlaminar shear modulus.

From the DMA thermograms obtained (figure 6), the glass transition temperature was analysed as the peak ($T_{g-peak}$) of loss tangent (tan $\delta$). A similar result was obtained for the different laminates taking into account the standard deviation of the results (table 2). This behaviour has already been reported by other researchers in multiscale materials [35, 48]. Adding functionalized nanoreinforcements into epoxy matrices could activate several mechanisms that might contribute to either decrease or increase the $T_g$ simultaneously [49]. In this case, the GRMs are not functionalized but two of them (GO and rGO) present a high oxygen content that could interact with the epoxy resin leading to a non-stoichiometric balance in the resin and creating covalent bonds between it and the nanoparticles, which could decrease and increase the $T_g$ respectively. In the case of the GNPs, with very low oxygen content, the two opposite effects could be related to the presence of agglomerates acting as defects and the steric hindrance that could reduce the matrix mobility, thus increasing the $T_g$ (also could be applied to GO and rGO). The shape of the storage modulus and tan $\delta$ curves was similar for all the laminates, indicating that the polymer network formed is the same.

The TGA thermograms obtained (figure S9) showed similar shape, indicating no significant differences in the decomposition process. The first weight decrease (onset between 250 °C and 400 °C) was analysed and considered the beginning of the thermal degradation process. All the laminates presented a similar decomposition temperature ($T_d$) with an increase of ~1 °C for the CFRP-GO and CFRP-rGO laminates (table 2), negligible considering the standard deviation of the results. Authors stated that an increment in the $T_d$ can be caused by the tortuous path effect, based on the barrier effect triggered by the nanoparticles, in which the entrance of oxygen is limited and the elimination of the volatile products delayed [50]. However, this effect depends on the quantity, exfoliation and dispersion of the GRMs into the resin [51], resulting in studies in which this effect is negligible [52] as in our case, where no effects can be appreciated due to the low number of nanoparticles (0.68 wt.%) inside the composite samples.

### 3.2.3. Mechanical properties of the laminates

A broad mechanical characterization of the different multiscale laminates was performed to analyse the effect of the GRMs in its mechanical behaviour. Tensile, flexural, interlaminar shear, in-plane shear and Mode I interlaminar facture toughness properties have been evaluated. Details about the calculation methods used are given in the SI.

Regarding the tensile tests, hardly any differences could be observed between the results obtained for the multiscale laminates compared to those obtained for the CFRP laminate (figures 7(a) and (b)). It should be noted that the tensile tests were performed in the 0° direction, which means that these properties are highly dependent on the CF behaviour. Considering that the CFs constitute 56% of the volume of the samples, the contribution of the GRMs (0.68 wt.% of the sample) is very small, which could explain the behaviour of CFRP-GO and CFRP-rGO laminates. However, when looking at the CFRP-GNP laminate, a decrease of 11% in the tensile strength is observed. In addition, the deviation obtained in this material (20%) is considerable higher than the deviation obtained in the other laminates (4%–5%), indicating that some coupons broke prematurely (figure S10). This is could be caused by the aggregates observed in figures 3 and S8(b), which act as crack nucleators, prompting an early failure. Several authors report analogous behaviours; Yokozeki *et al* obtained similar results in tensile tests for CF-epoxy composites manufactured from prepregs with 0% to 5% of cup-stacked CNTs [37]; Ashori *et al* observed a detriment to the tensile properties when adding more than 0.3 wt.% of functionalized GO [53]; and Gojny *et al* did not perceive either a difference in the tensile strength or Young's modulus of CFRP with a 0.1% of CNT both in the 0° and 90° directions [54]. Representative failure modes of each laminate can be observed in figure S11 where no differences are appreciated.

Flexural properties, in the same way, are sensitive primarily to volume fraction and mechanical properties of the CFs [55]. Results obtained are presented in figures 7(a) and (b), showing that multiscale laminates behave similarly to the non-nanoreinforced CFRP ones. No differences were appreciated neither between their failure modes (figure S12). Similar results were obtained by Quin *et al* for CF-epoxy composites manufactured by a prepreg lay-up with non-coated and GNP-coated CFs [55]; also by Siddiqui *et al* and Inam *et al* when adding functionalized CNT to CF reinforced epoxy composites [35, 48] and by Kamar *et al* when adding more than 1 wt. % of GNPs [56]. In this case, the deviation obtained in flexural strength in the CFRP-GNP laminate is within normal (4%) and no detriment in strength was observed. This difference with respect to the results obtained in tensile tests for CFRP-GNP laminates (despite of both being dependent on the CF behaviour), can be explained by the difference in stress distribution in both tests. In tensile tests, the whole coupon is subjected to a uniform stress condition; while in flexural test the distribution of stress is not constant, being the maximum in the section immediately under the roller (figure S13). Considering this, the probability of finding a defect (aggregate) that lead to a premature failure of the coupon is considerably higher when subjecting a coupon to a tensile test, being therefore the tensile test more sensitive to a poor dispersion of the nanoreinforcements.

On the other hand, the interlaminar shear strength (ILSS) and modulus have been obtained (figures 7(c) and (d)). Interlaminar shear tests in CFRP laminates usually prompts failure at the interply, as it is a resin rich region, where the matrix-fibre interactions and matrix properties play a key role [15, 57]. In this case, CFRP-rGO laminates presented 3% of improvement in strength compared to the CFRP laminate, what might indicate an enhancement in the interfacial adhesion [55, 58, 59]. The fact that the improvement is just seen in the CFRP-rGO could be related to the higher percentage of carboxyl groups (table 1), as the graphene-C=O/epoxy system present higher interface attraction than graphene/epoxy or graphene-OH/epoxy system [60]. The fact that CFRP-GNP laminate does not show a decrement in strength due to the presence of agglomeration could be due to the low GRM content, poor dispersion and the stress distribution in the tests (like the flexural test, figure S13) as explained above. This could explain as well the shy increment observed for CFRP-rGO and the 'no effect' in CFRP-GO laminate. Figure S14 shows the representative failure modes, where a secondary failure mode by flexion was observed for some coupons, mainly in CFRP-GNP laminate, which could indicate the presence of aggregates as observed in figures 3 and S7(b).

Regarding the ±45° tensile tests, the CFRP-GO laminate showed an increase of 6.5% in in-plane shear strength (IPSS) and 2.6% in modulus with respect to the CFRP laminate (figures 7(c) and (d)). These are complex tests in which matrix-dominated properties are evaluated [61, 62]. The affinity of the hydroxyl groups of the GO and the polar groups of the epoxy has been proved to improve the dispersion of the nanoreinforcements [19, 21], what might have led to an enhancement of the stress transfer among the composite [21]. On the other hand, CFRP-GNP and CFRP-rGO laminates present penalties of 9% and 12% in IPSS, respectively. As explained before, this test (tensile) entails a uniform stress distribution that makes it sensitive to the presence of aggregates (figure S13), indicating that GNP and rGO could be aggregated or bad distributed (GRM poor- and rich-regions) as seen in figures 3, 5 and S7(b) and S7(d).

Fracture surfaces of these coupons were already shown in section 3.2., where a difference in the fracture morphology between the CFRP (figure S7(a)) and the multiscale laminates (figures 3–5 and S7(b)–(d)) was observed. Figure S8(a) shows a slight hackle pattern in the resin indicative of mixed mode fracture and some clean fibres can be seen. On the other hand, a rougher surface is appreciated in the CFRP-GNP broken coupon (figures 3 and S7(b)) were GNP agglomerates were observed. Figure S8(c) showed a similar mixed failure mode to that observed for the CFRP (Figure S7(a)), however, the roughness found within the fibre prints (indicated in black circles) could be indicative of a good matrix/fibre interaction [23, 53, 63].

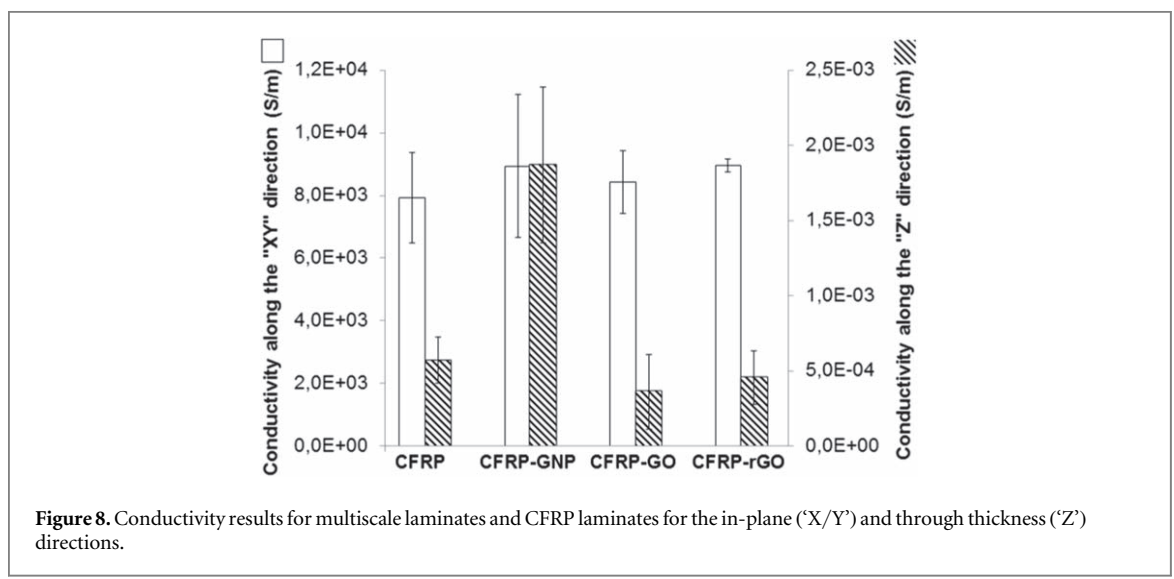

**Figure 8.** Conductivity results for multiscale laminates and CFRP laminates for the in-plane ('X/Y') and through thickness ('Z') directions.

**Table 3.** Maximum load, strain energy release rate at propagation point ($G_{IC, prop—MBT}$) and critical strain energy release rate ($G_{IC}$) results obtained in fracture toughness tests for CFRP and multiscale laminates.

|  | CFRP | CFRP-GNP | CFRP-GO | CFRP-rGO |
|---|---|---|---|---|
| Max Load (N) | 56 ± 3 | 54 ± 5 | 56 ± 3 | 59 ± 2 |
| $G_{IC, prop—MBT}$ (J m$^{-2}$) | 346 ± 35 | 309 ± 9 | 366 ± 17 | 376 ± 21 |
| $G_{IC}$ (J m$^{-2}$) | 210 ± 13 | 176 ± 8 | 204 ± 8 | 208 ± 14 |

Moreover, to characterize the delamination behaviour of the multiscale composites, mode I interlaminar fracture toughness tests were performed. The fracture toughness ($G_{IC}$), is the parameter used to measure the delamination resistance [13] which, in this case, has been calculated by the Area Method (area inside the curve divided by cracked area) [64]. In addition, the fracture toughness corresponding to the last crack propagation point -corresponding to a crack length around 62 mm- ($G_{IC, prop-MBT}$) has been calculated with the Modified Beam Theory (MBT) method, according to the ASTM D5528-01 standard. The results, together with the maximum load reached for each laminate, can be seen in table 3. Regarding the CFRP-GNP laminates, they present a decrease of 16% and 11% for both $G_{IC}$ values and 4% in maximum load achieved. These results are in accordance with the previously described (flexural, tensile, IPS), where the GNPs seem to be agglomerated and poor distributed, acting as paths for the cracks to propagate effortlessly [36]. Otherwise, CFRP-GO and CFRP-rGO laminates present an enhancement of the fracture toughness of 6 and 9% correspondingly when analysing the last propagation point. In addition, the maximum load achieved in CFRP-rGO laminate is 4.5% higher than the non-charged CFRP laminate. The observed enhancement could be related to a crack deflection mechanism, where the nanoreinforcements act as crack arresters [65, 66]. This could explain the increment in the maximum load reached at the beginning of the tip propagation, where the epoxy resin has not still established the full interaction with the CFs [36]. In addition, when looking at the representative *load* versus *displacement* curves (figure S16), a change in the curve roughness respect to the CFRP laminates could be appreciated, what might indicate unstable crack growth due to the presence of GRMs [48]. This behaviour is accentuated in the CFRP-GNP laminate, in which a stick-slip failure mode is found (figure S16), what may be caused by the presence of agglomerates. Representative pictures of the failure modes are shown in figure S17.

### 3.2.4. Electrical conductivity of the laminates

Considering that these materials have a linear electrical behaviour, the slope of every Intensity-Voltage curve was calculated by the least-squares method, obtaining the electrical resistance of every panel by the Ohm's Law. The results displayed in figure 8 indicate the electrical conductivity for each material, calculated considering the sample dimensions. The electrical conductivity along the 'X/Y' direction (in-plane) does not present significant changes between the materials. This behaviour is expected considering that the electrical conductivity along the 'X/Y' direction is dominated by the oriented CFs (with $10^3$–$10^6$ S m$^{-1}$) [67] that create conductive pathways along the plane, explaining the exhibited anisotropy of this property between 'X/Y' and 'Z' (trough the thickness) directions as well. Regarding the results obtained in the along the 'Z' direction, the CFRP-GNP laminates show an increment of 227%, which seems to indicate that GNPs percolate forming conductive paths between fibres and nanoparticles [52, 68]. On the other hand, the CFRP-GO presents a reduction of 37% and CFRP-rGO of 21%, explained by the modification of the GRMs, which implies structural changes that cause a detriment in the electrical conductivity of the GRMs, with the laminate containing more oxygen being more insulating.

## 4. Conclusions

Multiscale CFRPs, with epoxy matrix and aeronautical grade carbon fibre, containing three different GRMs have been successfully manufactured by prepreg hand lay-up and autoclave curing using standard procedures implemented in the industry in every step of the procedure: from the production of GRMs to the final laminates. In addition, the multiscale laminates manufactured showed the quality was as good as the laminate without GRMs, as seen by C-Scans, optical microscopy and acid digestion tests where negligible void content and no delamination was appreciated.

A complete characterization has been performed to evaluate the multiscale laminates and different behaviours could be observed. In general terms, the multiscale laminates containing GNPs have shown detriments regarding to the mechanical performance in comparison to the non-filled CFRP laminate, probably caused by a bad dispersion of the nanoparticles causing aggregates. Otherwise, electrical conductivity along the thickness direction is improved for CFRP-GNP laminates a 227%. On the other hand, the CFRP-rGO laminates seem to have lightly enhanced the matrix-fibre interface but probably a bad distribution of the nanoparticles impeded obtaining higher improvements. Finally, GO seem to be better integrated resulting in an improvement of the matrix and therefore in-plane shear properties and delamination resistance.

The full industrial manufacturing of the multiscale composites provides a step forward to the application of this technology into the industry. However, the discrete results obtained indicate that the dispersion strategies of GRMs within prepregs have to be optimized and work need to be done in order to understand how the pre-impregnation technique influences this dispersion.

## Acknowledgments

The research leading to this study has received funding from the European Union's Horizon 2020 research and innovation programme under Graphene Flagship, grant agreements 696656, 785219 and 881603. The authors want to acknowledge Delta-Tech for the support and the production of the prepregs, Airbus for their support for the research and development projects, Instituto de Tecnologías Químicas Emergentes de La Rioja-INTERQUIMICA for the XPS analysis of the GRMs and FIDAMC's workshop and laboratory staff for their help with the manufacturing and testing of the laminates.

## ORCID iDs

Verónica Rodríguez-García  https://orcid.org/0000-0002-3662-5809
Julio Gómez  https://orcid.org/0000-0002-6749-4354
María R Gude  https://orcid.org/0000-0003-4623-8519

## References

[1] Bacon R and Moses C T 1986 *High Performance Polymers: their Origin and Development* (Dordrecht: Springer)
[2] Hyer M W 2009 *Stress Analysis of Fiber-Reinforced Composite Materials* (Lancaster, Pennsylvania: DEStech Publications, Inc.)
[3] Launey M E and Ritchie R O 2009 On the fracture toughness of advanced materials *Adv. Mater.* **21** 2103–10
[4] Tang Y, Ye L, Zhang Z and Friedrich K 2013 Interlaminar fracture toughness and CAI strength of fibre-reinforced composites with nanoparticles—a review *Compos. Sci. Technol.* **86** 26–37
[5] Tong L, Mouritz A P and Bannister M K 2002 *3D Fibre Reinforced Polymer Composites* (Amsterdam: Elsevier)
[6] Kostopoulos V, Masouras A, Baltopoulos A, Vavouliotis A, Sotiriadis G and Pambaguian L 2017 A critical review of nanotechnologies for composite aerospace structures *CEAS Space J.* **9** 35–57
[7] Guo M, Yi X, Liu G and Liu L 2014 Simultaneously increasing the electrical conductivity and fracture toughness of carbon-fiber composites by using silver nanowires-loaded interleaves *Compos. Sci. Technol.* **97** 27–33
[8] Guo M, Yi X, Rudd C and Liu X 2019 Preparation of highly electrically carbon-fiber composites with high interlaminar fracture toughness by using silver-plated interleaves *Compos. Sci. Technol.* **176** 29–36
[9] Kuwata M and Hogg P J 2011 Interlaminar toughness of interleaved CFRP using non-woven veils: I. Mode-I testing *Compos. Part A: Appl. Sci. Manuf.* **42** 1551–9
[10] Quan D, Bologna F, Scarselli G, Ivankovic A and Murphy N 2020 Interlaminar fracture toughness of aerospace-grade carbon fibre reinforced plastics interleaved with thermoplastic veils *Compos. Part A: Appl. Sci. Manuf.* **128** 105642
[11] Latko P, Ruminski W and Boczkowska A 2015 Carbon nanotubes-doped veils *Compos. Struct.* **134** 52–9
[12] Njuguna J, Pielichowski K and Alcock J R 2007 Epoxy-based fibre reinforced nanocomposites *Adv. Eng. Mater.* **9** 835–47

[13] Wetzel B, Rosso P, Haupert F and Friedrich K 2006 Epoxy nanocomposites—fracture and toughening mechanisms *Eng. Fract. Mech.* **73** 2375–98
[14] Dikshit V, Bhudolia S and Joshi S 2017 Multiscale polymer composites: a review of the interlaminar fracture toughness improvement *Fibers* **5** 38
[15] Valorosi F *et al* 2020 Graphene and related materials in hierarchical fiber composites: production techniques and key industrial benefits *Compos. Sci. Technol.* **185** 107848
[16] Geim A K and Novoselov K S 2007 The rise of graphene *Nature Mater.* **6** 183–91
[17] Papageorgiou D G, Li Z, Liu M, Kinloch I A and Young R J 2020 Mechanisms of mechanical reinforcement by graphene and carbon nanotubes in polymer nanocomposites *Nanoscale* **12** 2228–67
[18] Li B and Zhong W-H 2011 Review on polymer/graphite nanoplatelet nanocomposites *J. Mater Sci.* **46** 5595–614
[19] Young R J, Kinloch I A, Gong L and Novoselov K S 2012 The mechanics of graphene nanocomposites: a review *Compos. Sci. Technol.* **72** 1459–76
[20] Konios D, Stylianakis M M, Stratakis E and Kymakis E 2014 Dispersion behaviour of graphene oxide and reduced graphene oxide *J. Colloid Interface Sci.* **430** 108–12
[21] Li Z, Young R J, Wang R, Yang F, Hao L, Jiao W and Liu W 2013 The role of functional groups on graphene oxide in epoxy nanocomposites *Polymer* **54** 5821–9
[22] Ahmadi-Moghadam B, Sharafimasooleh M, Shadlou S and Taheri F 2015 Effect of functionalization of graphene nanoplatelets on the mechanical response of graphene/epoxy composites *Mater. Des. (1980–2015)* **66** 142–9
[23] Chandrasekaran S, Sato N, Tölle F, Mülhaupt R, Fiedler B and Schulte K 2014 Fracture toughness and failure mechanism of graphene based epoxy composites *Compos. Sci. Technol.* **97** 90–9
[24] Choi W, Lahiri I, Seelaboyina R and Kang Y S 2010 Synthesis of graphene and its applications: a review *Crit. Rev. Solid State Mater. Sci.* **35** 52–71
[25] Paton K R *et al* 2014 Scalable production of large quantities of defect-free few-layer graphene by shear exfoliation in liquids *Nature Mater.* **13** 624–30
[26] Martin-Gallego M, Bernal M M, Hernandez M, Verdejo R and Lopez-Manchado M A 2013 Comparison of filler percolation and mechanical properties in graphene and carbon nanotubes filled epoxy nanocomposites *Eur. Polym. J.* **49** 1347–53
[27] Rafiee M, Nitzsche F, Laliberte J, Hind S, Robitaille F and Labrosse M R 2019 Thermal properties of doubly reinforced fiberglass/epoxy composites with graphene nanoplatelets, graphene oxide and reduced-graphene oxide *Compos. Part B Eng.* **164** 1–9
[28] Yue L, Pircheraghi G, Monemian S A and Manas-Zloczower I 2014 Epoxy composites with carbon nanotubes and graphene nanoplatelets—dispersion and synergy effects *Carbon* **78** 268–78
[29] Reia da Costa E F, Skordos A A, Partridge I K and Rezai A 2012 RTM processing and electrical performance of carbon nanotube modified epoxy/fibre composites *Compos. Part A: Appl. Sci. Manuf.* **43** 593–602
[30] Zhang H, Liu Y, Huo S, Briscoe J, Tu W, Picot O T, Rezai A, Bilotti E and Peijs T 2017 Filtration effects of graphene nanoplatelets in resin infusion processes: Problems and possible solutions *Compos. Sci. Technol.* **139** 138–45
[31] Jiménez-Suárez A, Campo M, Prolongo S G, Sánchez M and Ureña A 2016 Effect of filtration in functionalized and non-functionalized CNTs and surface modification of fibers as an effective alternative approach *Compos. Part B Eng.* **94** 286–91
[32] Hung P, Lau K, Fox B, Hameed N, Jia B and Lee J H 2019 Effect of graphene oxide concentration on the flexural properties of CFRP at low temperature *Carbon* **152** 556–64
[33] Islam A B M I and Kelkar A D 2017 Prospects and challenges of nanomaterial engineered prepregs for improving interlaminar properties of laminated composites—a review *MRC* **7** 102–8
[34] Siddiqui N A, Khan S U, Ma P C, Li C Y and Kim J-K 2011 Manufacturing and characterization of carbon fibre/epoxy composite prepregs containing carbon nanotubes *Compos. Part A: Appl. Sci. Manuf.* **42** 1412–20
[35] Siddiqui N A, Khan S U and Kim J-K 2013 Experimental torsional shear properties of carbon fiber reinforced epoxy composites containing carbon nanotubes *Compos. Struct.* **104** 230–8
[36] Joshi S C and Dikshit V 2012 Enhancing interlaminar fracture characteristics of woven CFRP prepreg composites through CNT dispersion *J. Compos. Mater.* **46** 665–75
[37] Yokozeki T, Iwahori Y, Ishiwata S and Enomoto K 2007 Mechanical properties of CFRP laminates manufactured from unidirectional prepregs using CSCNT-dispersed epoxy *Compos. Part A: Appl. Sci. Manuf.* **38** 2121–30
[38] Wang P-N, Hsieh T-H, Chiang C-L and Shen M-Y 2015 Synergetic effects of mechanical properties on graphene nanoplatelet and multiwalled carbon nanotube hybrids reinforced epoxy/carbon fiber composites *J. Nanomater.* **2015** 1–9
[39] Mannov E, Schmutzler H, Chandrasekaran S, Viets C, Buschhorn S, Tölle F, Mülhaupt R and Schulte K 2013 Improvement of compressive strength after impact in fibre reinforced polymer composites by matrix modification with thermally reduced graphene oxide *Compos. Sci. Technol.* **87** 36–41
[40] Ferrari A C *et al* 2015 Science and technology roadmap for graphene, related two-dimensional crystals, and hybrid systems *Nanoscale* **7** 4598–810
[41] Saini A 2014 EU Graphene Flagship project aims for technological breakthroughs: graphene-flagship.eu *MRS Bulletin* **5** 393–4
[42] Elmarakbi A, Karagiannidis P, Ciappa A, Innocente F, Galise F, Martorana B, Bertocchi F, Cristiano F, Villaro Ábalos E and Gómez J 2019 3-Phase hierarchical graphene-based epoxy nanocomposite laminates for automotive applications *J. Mater. Sci. Technol.* **35** 2169–77
[43] Gómez J, Villaro E, Navas A and Recio I 2017 Testing the influence of the temperature, RH and filler type and content on the universal power law for new reduced graphene oxide TPU composites *Mater. Res. Express.* **4** 105020
[44] Stoller M D, Park S, Zhu Y, An J and Ruoff R S 2008 Graphene-based ultracapacitors *Nano Lett.* **8** 3498–502
[45] Georgakilas V, Otyepka M, Bourlinos A B, Chandra V, Kim N, Kemp K C, Hobza P, Zboril R and Kim K S 2012 Functionalization of graphene: covalent and non-covalent approaches, derivatives and applications *Chem. Rev.* **112** 6156–214
[46] Kuilla T, Bhadra S, Yao D, Kim N H, Bose S and Lee J H 2010 Recent advances in graphene based polymer composites *Progr. Polym. Sci.* **35** 1350–75
[47] Wang F, Drzal L T, Qin Y and Huang Z 2016 Size effect of graphene nanoplatelets on the morphology and mechanical behavior of glass fiber/epoxy composites *J. Mater. Sci.* **51** 3337–48
[48] Inam F, Wong D W Y, Kuwata M and Peijs T 2010 Multiscale hybrid micro-nanocomposites based on carbon nanotubes and carbon fibers *J. Nanomater.* **2010** 1–12
[49] Shen J, Huang W, Wu L, Hu Y and Ye M 2007 The reinforcement role of different amino-functionalized multi-walled carbon nanotubes in epoxy nanocomposites *Compos. Sci. Technol.* **67** 3041–50

[50] Zhou T, Liu F, Suganuma K and Nagao S 2016 Use of graphene oxide in achieving high overall thermal properties of polymer for printed electronics *RSC Adv.* **6** 20621–8

[51] Prolongo S G, Jimenez-Suarez A, Moriche R and Ureña A 2013 *In situ* processing of epoxy composites reinforced with graphene nanoplatelets *Compos. Sci. Technol.* **86** 185–91

[52] Alhassan S M, Qutubuddin S, Schiraldi D A, Agag T and Ishida H 2013 Preparation and thermal properties of graphene oxide/main chain benzoxazine polymer *Eur. Polym. J.* **49** 3825–33

[53] Ashori A, Rahmani H and Bahrami R 2015 Preparation and characterization of functionalized graphene oxide/carbon fiber/epoxy nanocomposites *Polym. Test.* **48** 82–8

[54] Gojny F H, Wichmann M H G, Fiedler B, Bauhofer W and Schulte K 2005 Influence of nano-modification on the mechanical and electrical properties of conventional fibre-reinforced composites *Compos. Part A: Appl. Sci. Manuf.* **36** 1525–35

[55] Qin W, Vautard F, Drzal L T and Yu J 2015 Mechanical and electrical properties of carbon fiber composites with incorporation of graphene nanoplatelets at the fiber–matrix interphase *Compos. Part B Eng.* **69** 335–41

[56] Kamar N T, Hossain M M, Khomenko A, Haq M, Drzal L T and Loos A 2015 Interlaminar reinforcement of glass fiber/epoxy composites with graphene nanoplatelets *Compos. Part A: Appl. Sci. Manuf.* **70** 82–92

[57] Tan W, Naya F, Yang L, Chang T, Falzon B G, Zhan L, Molina-Aldareguía J M, González C and Llorca J 2018 The role of interfacial properties on the intralaminar and interlaminar damage behaviour of unidirectional composite laminates: experimental characterization and multiscale modelling *Compos. Part B Eng.* **138** 206–21

[58] Zhang X, Fan X, Yan C, Li H, Zhu Y, Li X and Yu L 2012 Interfacial microstructure and properties of carbon fiber composites modified with graphene oxide *ACS Appl. Mater. Interfaces.* **4** 1543–52

[59] Lee M-W, Wang T-Y and Tsai J-L 2016 Characterizing the interfacial shear strength of graphite/epoxy composites containing functionalized graphene *Compos. Part B Eng.* **98** 308–13

[60] Melro L S and Jensen L R 2020 Interfacial characterization of functionalized graphene-epoxy composites *J. Comp. Mater.* **54** 703–10

[61] Cho J, Chen J Y and Daniel I M 2007 Mechanical enhancement of carbon fiber/epoxy composites by graphite nanoplatelet reinforcement *Scripta Materialia* **56** 685–8

[62] Godara A, Mezzo L, Luizi F, Warrier A, Lomov S V, van Vuure A W, Gorbatikh L, Moldenaers P and Verpoest I 2009 Influence of carbon nanotube reinforcement on the processing and the mechanical behaviour of carbon fiber/epoxy composites *Carbon* **47** 2914–23

[63] Garg M, Sharma S and Mehta R 2015 Pristine and amino functionalized carbon nanotubes reinforced glass fiber epoxy composites *Compos. Part A: Appl. Sci. Manuf.* **76** 92–101

[64] Whitney J M, Browning C E and Hoogsteden W 1981 A double cantilever beam test for characterizing mode I delamination of composite materials *J. Reinf. Plast. Compos.* **1** 297–313

[65] Bandyopadhyay S 1990 Review of the microscopic and macroscopic aspects of fracture of unmodified and modified epoxy resins *Mater. Sci. Eng.* A **125** 157–84

[66] Fiedler B, Gojny F H, Wichmann M H G, Nolte M C M and Schulte K 2006 Fundamental aspects of nano-reinforced composites *Compos. Sci. Technol.* **66** 3115–25

[67] Banerjee P and Schmidt J L 2013 Electrical conductivity modeling and validation in unidirectional carbon fiber reinforced polymer composites *Proc. of the COMSOL Conf.*

[68] Hashemi R and Weng G J 2016 A theoretical treatment of graphene nanocomposites with percolation threshold, tunneling-assisted conductivity and microcapacitor effect in AC and DC electrical settings *Carbon* **96** 474–90